\documentclass[sigconf]{acmart}
\renewcommand\footnotetextcopyrightpermission[1]{} 
\AtBeginDocument{%
  }

\usepackage{pifont}          
\usepackage{xcolor}          
\usepackage{colortbl}        
\usepackage{multirow}        
\usepackage{cite}

\usepackage{amsmath,amssymb,amsfonts}
\usepackage{algorithm}    
\usepackage{url}
\usepackage{booktabs}
\usepackage{xurl}
\usepackage{caption}
\usepackage{subcaption}
\usepackage{multicol}
\usepackage{algpseudocode}
\usepackage{graphicx}
\usepackage{textcomp}
\usepackage{xcolor}
\usepackage{color}
\usepackage{listings}
\usepackage{xspace}
\usepackage{booktabs}
\usepackage{array}
\usepackage{enumitem}
\usepackage{tcolorbox}
\usepackage{array}

\newcommand{\method}{\textsc{JSProtect}\xspace}

\setcopyright{acmlicensed}
\copyrightyear{2018}
\acmYear{2018}
\acmDOI{XXXXXXX.XXXXXXX}
\acmConference[Conference acronym 'XX]{Make sure to enter the correct
  conference title from your rights confirmation email}{June 03--05,
  2018}{Woodstock, NY}
\acmISBN{978-1-4503-XXXX-X/2018/06}

\begin{document}

\title{\method: A Scalable Obfuscation Framework for Mini-Games in WeChat}


\author{%
\begin{tabular}[t]{c}
Zhihao Li\textsuperscript{1}$^{\dagger}$, \and Chaozheng Wang\textsuperscript{1}\textsuperscript{2}$^{\dagger}$, \and Zongjie Li\textsuperscript{3}$^{*}$, \and Xinyong Peng\textsuperscript{1}, \and Zelin Su\textsuperscript{1}, \and Qun Xia\textsuperscript{1}  \\
Haochuan Lu\textsuperscript{1}, \and Ting Xiong\textsuperscript{1}, \and Man Ho Lam\textsuperscript{2}, \and Shuzheng Gao\textsuperscript{2}, \and Yuchong Xie\textsuperscript{3}\\ 
Cuiyun Gao\textsuperscript{2}, \and Shuai Wang\textsuperscript{3}, \and Yuetang Deng\textsuperscript{1}, \and Huafeng Ma\textsuperscript{1}\\
\footnotesize $^{\dagger}$These authors contributed equally to this work. $^{*}$Corresponding author.
\end{tabular}
}

\affiliation{%
  $^1$Tencent Inc., $^2$The Chinese University of Hong Kong, $^3$Hong Kong University of Science and Technology
  \country{China}
}

\renewcommand{\shortauthors}{Trovato et al.}

\begin{abstract}
The WeChat mini-game ecosystem faces rampant intellectual property theft to other platforms via secondary development, yet existing JavaScript obfuscation tools are ill-equipped for large-scale applications, suffering from prohibitive processing times, severe runtime performance degradation, and unsustainable code size inflation. This paper introduces \method, a high-throughput parallelized obfuscation framework designed to overcome these fundamental limitations. At the core of our framework is the Parallel-Aware Scope Analysis (PASA) algorithm, which enables two key optimizations: independent code partitioning for multi-core processing and independent namespace management that aggressively reuses short identifiers to combat code bloat. Our evaluation demonstrates that \method processes 20MB codebases in minutes, maintaining 100\% semantic equivalence while controlling code size inflation to as low as 20\% compared to over 1,000\% with baseline tools. Furthermore, it preserves near-native runtime performance and provides superior security effectiveness against both static analysis tools and large language models. This work presents a new paradigm for industrial-scale JavaScript protection that effectively balances robust security with high performance and scalability.
\end{abstract}

\received{20 February 2007}
\received[revised]{12 March 2009}
\received[accepted]{5 June 2009}

\maketitle

\section{Introduction}
The proliferation of smartphone technology has fundamentally transformed the mobile gaming landscape, driving unprecedented growth in this sector over the past decade. According to industry analytics, global mobile game revenue is projected to reach \$135 billion across app stores~\citep{data}. This immense profitability has not only fostered a burgeoning ecosystem of game development but has also led to the proliferation of various mini-game platforms. These platforms distinguish themselves from traditional mobile game applications by embedding lightweight games directly within existing mobile applications, thereby eliminating the need for separate downloads or installations. This ``plug-and-play'' functionality substantially enhances user engagement and has attracted a vast user base, alongside a rapidly expanding community of developers and platform providers. A prime example of such a successful ecosystem is the Tencent \textit{WeChat} mini-game platform, which has garnered hundreds of thousands of developers and hosts hundreds of thousands of mini-games, serving a user base exceeding one billion and boasting 500 million monthly active users~\citep{wang2023unified}.

The rapid expansion of mini-game ecosystems, such as WeChat's, has also introduced significant security challenges, particularly in the form of game plagiarism and intellectual property theft. In prior work~\citep{li2025jsidentify, xia2020jsidentify}, researchers have explored aspects of security protection for WeChat mini-games, including code plagiarism detection mechanisms that identify duplicated code submissions within the platform. While these detection approaches effectively prevent plagiarized games from appearing on WeChat itself, they fall short in addressing a more insidious form of exploitation: code porting. This involves adversaries stealing the source code, performing secondary development (e.g., replacing WeChat-specific libraries, making minor functional modifications, or swapping assets), and then deploying the modified game on alternative platforms. Such practices severely undermine developers' intellectual property rights, erode competitive advantages, and result in substantial economic losses.

To mitigate the risks of secondary development and code porting, code obfuscation emerges as a widely adopted defensive technique, which transforms readable source code into an equivalent but semantically opaque form, thereby increasing the difficulty of reverse engineering. 
Existing JavaScript obfuscation tools employ various transformations to disrupt code analysis by attackers, with some advanced academic approaches leveraging WebAssembly translation for enhanced protection~\citep{romano2022wobfuscator}.
However, existing JavaScript obfuscation tools face three critical challenges when applied to WeChat Mini-Game scenarios: (1) \textbf{Scalability Crisis}: processing time and memory consumption escalate superlinearly with code size, with tools requiring over 13 hours to process 20 MB codebases while consuming up to 22 GB memory; (2) \textbf{Runtime Performance Degradation}: obfuscation introduces substantial computational overhead through additional branches, jumps, and repeated calculations, with some tools degrading frame rates from 60 FPS to as low as 1 FPS and increasing execution time by over 80$\times$; and (3) \textbf{Severe Code Size Inflation}: traditional techniques cause dramatic code expansion (10-20$\times$ inflation), transforming 20 MB games into 200 MB+ outputs through loop unrolling, dead code injection, and inefficient variable naming strategies that exhaust short identifiers and resort to lengthy random hexadecimal strings for millions of variables.

This paper introduces a parallelized obfuscation framework tailored for large-scale JavaScript applications, with core innovations centered around deep scope analysis that enables multiple optimization strategies to address these three fundamental challenges simultaneously. Our key insight is that comprehensive scope analysis serves as the foundation for both aggressive code size reduction and scalable parallel processing, while custom performance optimizations preserve runtime efficiency.
At the core of our approach is a sophisticated \textbf{Parallel-Aware Scope Analysis (PASA)} algorithm that constructs detailed scope trees and identifies cross-scope dependencies, enabling three critical capabilities that directly address our key challenges: (1) independent namespace management that solves the code size inflation problem by treating each function scope as an isolated namespace, allowing aggressive reuse of short identifiers across disjoint scopes rather than exhausting global namespaces with lengthy variable names; (2) independent code partitioning that addresses the scalability crisis by identifying semantically independent code units suitable for parallel processing, eliminating the dependencies that force traditional tools into single-threaded execution; and (3) performance-aware obfuscation strategies that preserve runtime efficiency through optimized transformations and memoization techniques, ensuring minimal computational overhead while maintaining strong security properties.

Our comprehensive evaluation demonstrates that \method achieves superior performance across all key metrics compared to existing baselines. In terms of scalability, \method exhibits the most robust performance when facing increasing source code sizes, reducing processing time by 84\% and memory consumption by 67\% compared to the best commercial tools. The obfuscated code generated by our framework also achieves optimal runtime performance, delivering the fastest execution speeds and lowest memory overhead in both third-party library benchmarks and mini-game applications. Furthermore, our approach demonstrates the highest security effectiveness, producing obfuscated code with maximum complexity that proves most resistant to analysis by both static analysis tools and large language models, thereby providing the strongest protection against reverse engineering attempts.  Currently, approximately 10,000 active mini-games utilize our obfuscation methodology, running on more than 100 million user devices. Through comprehensive monitoring and analysis of game submissions across multiple mini-game platforms, we observed that instances of game plagiarism have decreased by \textbf{91\%} over the past year compared to pre-deployment statistics.

This paper makes the following key technical contributions:

\begin{itemize}
\item \textbf{Parallel-Aware Scope Analysis (PASA) Algorithm}: We propose a novel scope analysis algorithm that enables parallel obfuscation by identifying scope independence and dependency boundaries, serving as the foundation for both scalable processing and aggressive namespace optimization.

\item \textbf{High-Throughput Parallel Obfuscation Framework}: We design \method, a parallelized obfuscation system that processes large-scale JavaScript codebases (20MB+) in minutes through independent code partitioning, achieving near-linear scalability with multi-core architectures.

\item \textbf{Scope-Aware Namespace Management}: We develop an independent namespace strategy that enables aggressive reuse of short identifiers across disjoint scopes, reducing code size inflation from over 1,000\% (baseline tools) to as low as 20\% while preserving obfuscation effectiveness.

\item \textbf{Performance-Preserving Obfuscation Techniques}: We introduce performance-aware transformations, including optimized property access and memoization strategies that maintain near-native runtime performance while providing robust security against both static analysis and LLM-based reverse engineering.
\end{itemize}
\section{Background and Problem Definition}

\subsection{Threat Model and Attack Scenarios}

In the WeChat Mini-Game ecosystem, we face diverse adversaries with varying capabilities and motivations. Code pirates represent the most prevalent threat, systematically copying entire games to alternative platforms to circumvent WeChat's revenue-sharing mechanisms while claiming authorship. Cheat developers analyze game logic and data structures to create tools that modify game behavior, undermining competitive balance and monetization systems. Algorithm thieves target core computational components such as AI systems, physics engines, and economic models for integration into competing products. Reverse engineers conduct deep structural analysis to extract business secrets and implementation details, potentially exposing intellectual property and competitive advantages~\citep{brezinski2023metamorphic,schloegel2022loki}.

These threat actors employ increasingly sophisticated techniques, including automated static analysis, dynamic debugging, and machine learning-based pattern recognition. The distributed nature of JavaScript execution in web browsers provides attackers with complete access to source code, making traditional compilation-based protection mechanisms ineffective~\citep{maiorca2015stealth}. Furthermore, the open-source ecosystem surrounding JavaScript development provides attackers with extensive tooling for code analysis and manipulation.

\subsection{JavaScript Obfuscation Problem Formalization}

\textbf{Definition 2.1} (JavaScript Program). A JavaScript program $\mathcal{P}$ is a tuple $\langle \mathcal{F}, \mathcal{D}, \mathcal{E} \rangle$ where:
\begin{itemize}
    \item $\mathcal{F} = \{f_1, f_2, \ldots, f_n\}$ is a set of source files
    \item $\mathcal{D}$ represents the dependency graph among files
    \item $\mathcal{E}$ denotes the execution environment constraints
\end{itemize}

\textbf{Definition 2.2} (Code Obfuscation). Given a JavaScript program $\mathcal{P}$, an obfuscation transformation $\Omega$ is a function:
$$\Omega: \mathcal{P} \rightarrow \mathcal{P}'$$
such that:
\begin{enumerate}
    \item \textbf{Semantic Preservation}: $\llbracket \mathcal{P} \rrbracket_{\mathcal{E}} = \llbracket \mathcal{P}' \rrbracket_{\mathcal{E}}$ (same operational semantics)
    \item \textbf{Security Enhancement}: $\text{Secu}(\mathcal{P}') > \text{Secu}(\mathcal{P})$
    \item \textbf{Performance Constraint}: $\text{RT}(\mathcal{P}') \leq (1+\epsilon) \cdot \text{RT}(\mathcal{P})$ for small $\epsilon$
\end{enumerate}
where $\llbracket \cdot \rrbracket_{\mathcal{E}}$ denotes operational semantics under environment $\mathcal{E}$.

\subsection{Empirical Analysis: Scalability Crisis}
\label{subsec:empirical}
Traditional JavaScript obfuscation encompasses identifier renaming, string encryption, control flow flattening, and dead code injection. However, these techniques exhibit fundamental scalability limitations when applied to industrial-scale applications. We conduct a preliminary evaluation of existing obfuscation tools, such as JS-Obfuscator~\citep{JavaScriptdeobfuscator} on WeChat Mini-Games of varying scales, revealing systematic failures at industrial scale.

\subsubsection{Experimental Setup}

We select representative WeChat Mini-Games ranging from 500KB to 20MB, encompassing various frameworks. Each game is processed using leading obfuscation tools: JS-obfuscator, JSCrambler, and VirBox. Measurements include processing time, output code size, memory consumption, and runtime performance.

\subsubsection{Quantitative Results}

Our empirical findings across different code sizes and obfuscation tools demonstrate a clear scalability crisis as project size increases.

\textbf{Observation 1} (Exponential Time Growth). For traditional obfuscation tools, processing time exhibits super-linear growth:
$$T_{\text{sequential}}(|\mathcal{P}|) = \Theta(|\mathcal{P}|^{\alpha})$$
where $\alpha \geq 2.3$ for JS-obfuscator when $|\mathcal{P}| > 5$ MB.

The data reveal dramatic time scaling issues. JS-obfuscator's processing time grows from 9.9 seconds for 500KB to 47,445 seconds (approximately 13 hours) for 20MB input. This represents a 4,792× time increase for only a 40× size increase, demonstrating highly super-linear scaling behavior. The processing time escalates from manageable durations to completely impractical timeframes that exceed typical continuous integration pipeline tolerances by several orders of magnitude.

\textbf{Observation 2} (Memory Explosion). Memory consumption follows:
$$M_{\text{sequential}}(|\mathcal{P}|) = \Theta(|\mathcal{P}|^{\beta})$$
where $\beta \approx 1.4$ for JS-obfuscator, causing substantial memory pressure for large projects.

Memory usage patterns demonstrate problematic scaling characteristics. JS-obfuscator's memory consumption grows from 263 MB for 500 KB input to 22,035 MB (approximately 22 GB) for 20 MB input. This represents an 84× memory increase for a 40× input size increase, indicating memory requirements that exceed typical development machine capabilities for large projects. This memory explosion stems from the need to maintain entire AST structures in memory simultaneously during the obfuscation process.

\textbf{Observation 3} (Severe Code Size Inflation). Traditional obfuscation causes substantial code expansion:
$$|\mathcal{P}'| \approx (10 \text{ to } 20) \cdot |\mathcal{P}|$$
leading to output sizes that present deployment challenges.

Code size inflation presents a critical limitation for WeChat Mini-Game deployment. JS-obfuscator consistently produces 10-20× code expansion across all tested project sizes. This massive inflation stems from multiple obfuscation techniques that inherently increase code volume. Dead code injection introduces substantial amounts of non-functional code to obscure program logic, while loop unrolling expands compact iterative structures into verbose sequential operations. Control flow flattening transforms simple conditional structures into complex state machines with extensive branching logic.

Variable name expansion represents another critical source of code inflation. Traditional obfuscation methods employ random hexadecimal strings or alphabetical sequences (a, b, ..., z, aa, ab, ..., az, ba, ...), which become severely problematic for large-scale mini-games containing millions of identifiers. As namespaces exhaust shorter combinations, variable names expand from single characters to lengthy strings such as "zyxwvut" or long random hexadecimal sequences. For complex Mini-Games with original code sizes reaching 20 MB, this combined effect can produce obfuscated output exceeding 200 MB, creating substantial challenges for code storage, network transmission, and runtime performance in the constrained Mini-Game environment.

\textbf{Observation 4} (Runtime Performance Degradation). Traditional obfuscation tools cause severe runtime performance degradation that renders games unplayable:

\[
\text{FPS}_{\text{obfuscated}} \approx (0.4 \sim 0.8) \cdot \text{FPS}_{\text{original}}
\]
The runtime performance impact represents a critical limitation for interactive applications. Our preliminary analysis of JS-obfuscator on representative mini-games reveals substantial frame rate degradation. This performance penalty stems from the computational overhead introduced by obfuscation transformations, including additional conditional branches for control flow flattening, repeated string decryption operations, and complex property access patterns that replace direct object member access with function calls or computed expressions. For game applications where smooth 60 FPS performance is essential for user experience, such degradation is unacceptable to players and can result in negative reviews, user abandonment, and commercial failure in the competitive mobile gaming market.

\subsection{Mini-Game Deployment Requirements}

WeChat Mini-Game deployment scenarios impose unique constraints on obfuscation tools. Instant build requirements demand completion of the entire pipeline from code submission to production deployment within minutes, as developers expect rapid iteration cycles typical of web development. Package size limitations enforced by the WeChat platform require strict control of code inflation, as games exceeding size limits face rejection or performance penalties. Diverse codebase characteristics spanning multiple JavaScript frameworks and libraries necessitate exceptional compatibility, as obfuscation failures can break critical game functionality.

The empirical findings above demonstrate that sequential obfuscation fundamentally cannot scale to modern JavaScript applications, necessitating a paradigm shift toward parallel processing architectures that can handle industrial-scale codebases efficiently.

\section{Methodology}
\label{sec:method}

\begin{figure*}[t]
    \centering
    \includegraphics[width=0.92\textwidth]{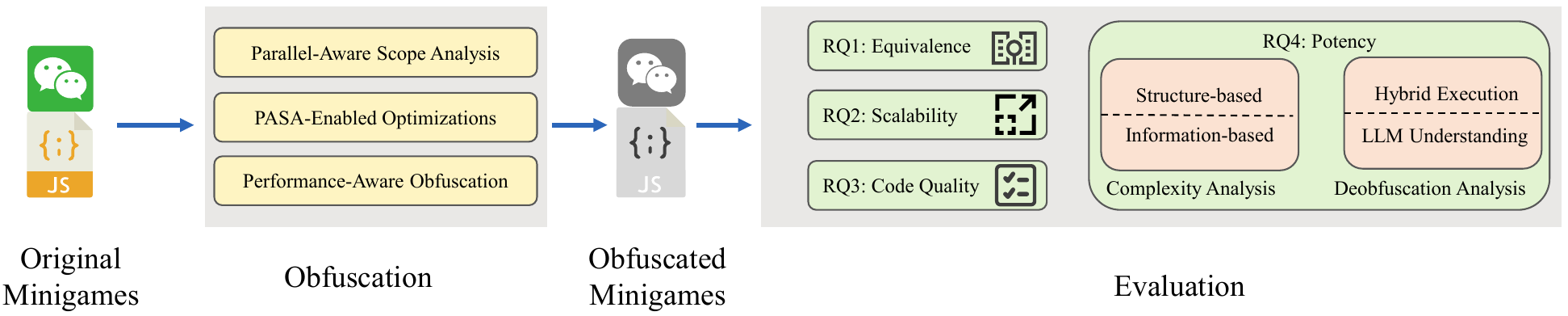}
    \vspace{-8pt}
    \caption{Overview of \method. }
    \label{fig:overview}
    \vspace{-8pt}
\end{figure*}

Our methodology, as shown in Figure \ref{fig:overview}, consists of three interconnected components that work together to achieve scalable, high-performance JavaScript obfuscation. The Parallel-Aware Scope Analysis (PASA) forms the foundation by identifying scope independence and dependency boundaries, which then enables two complementary optimization strategies: PASA-enabled optimizations for aggressive namespace management and independent code partitioning, and performance-aware obfuscation techniques that preserve runtime efficiency while maintaining strong security properties. 

\subsection{Parallel-Aware Scope Analysis (PASA)}

Scope analysis represents the foundational component of our entire obfuscation framework, as it enables both independent namespace management and parallel code processing. Traditional obfuscation tools perform global analysis that creates dependencies preventing parallelization, while our Parallel-Aware Scope Analysis (PASA) is specifically designed to support the parallel execution of subsequent optimization phases.

\subsubsection{Formal Scope Model}

\textbf{Definition 3.1} (Scope Tree). The scope structure of a JavaScript program is represented as a tree $\mathcal{S} = (V_S, E_S, \text{root})$ where:
\begin{itemize}
    \item $V_S$ is the set of scope nodes
    \item $E_S \subseteq V_S \times V_S$ represents lexical nesting relationships
    \item $\text{root} \in V_S$ denotes the global scope
\end{itemize}

Each scope node $s \in V_S$ is characterized by:
\begin{itemize}
    \item $\text{Declarations}(s)$: identifiers declared within scope $s$
    \item $\text{References}(s)$: identifiers referenced but not declared in $s$
    \item $\text{Type}(s) \in \{\text{function}, \text{block}, \text{module}, \text{global}\}$
    \item $\text{IsDynamic}(s) \in \{\text{true}, \text{false}\}$: presence of eval, with statements, etc.
\end{itemize}

\textbf{Definition 3.2} (Identifier Resolution). For identifier $x$ referenced in scope $s$, the resolution function $\rho$ is defined recursively:
$$\rho(x, s) = \begin{cases}
\langle s, x \rangle & \text{if } x \in \text{Declarations}(s) \\
\rho(x, \text{parent}(s)) & \text{if } s \neq \text{root} \\
\langle \text{global}, x \rangle & \text{otherwise}
\end{cases}$$

\subsubsection{Multi-Phase Parallel Analysis Architecture}

Our PASA algorithm addresses the fundamental challenge that traditional scope analysis requires global program understanding, making parallelization impossible. PASA introduces a multi-phase approach that enables independent processing of code segments while maintaining correctness, thereby serving as the foundation for all subsequent parallel optimizations.

The PASA algorithm operates through three coordinated phases to enable parallel processing. First, during the parsing phase, we simultaneously extract dependency information, identifying import/export relationships, global variable references, and cross-file function calls to create a dependency graph $G_D = (F, D)$ where $F$ represents files and $D$ represents dependencies between files, eliminating the need for separate dependency analysis passes and directly supporting parallel execution in subsequent phases. Second, files with minimal cross-dependencies are identified using graph partitioning algorithms that minimize edge cuts in $G_D$, allowing each partition to construct its internal scope trees independently since the majority of scope relationships are intra-file, which is crucial for enabling both parallel namespace management and independent code partitioning in later phases. Third, cross-file scope dependencies are resolved through a lightweight coordination phase that merges the independent scope trees and resolves global identifier bindings, processing only identifiers with cross-file visibility to reduce coordination overhead.

\textbf{PASA Output for Parallel Processing}. The PASA phase produces two critical data structures that enable subsequent parallel optimizations: (1) \textit{Scope Independence Map}: identifies which scopes can be processed independently without affecting others, directly enabling parallel namespace reuse; (2) \textit{Dependency Boundary Markers}: define the minimal coordination points required between parallel processing units, forming the basis for atomic code partitioning.

\subsection{PASA-Enabled Optimizations}

Building upon the scope independence information and dependency boundaries provided by PASA, we implement two complementary optimization strategies that directly address the scalability and code size inflation challenges. These optimizations are only possible due to the detailed scope analysis performed by PASA, which provides the necessary independence guarantees for safe parallel execution.

\subsubsection{Independent Namespace Management}
\label{subsec:namespacealgo}

The scope independence information from PASA enables aggressive namespace reuse by allowing different scopes to share short variable names without conflicts. This directly addresses the code size inflation problem that occurs when traditional obfuscation tools must use increasingly long variable names to avoid global naming conflicts.

\textbf{Safe Renaming Constraints}

\textbf{Definition 3.3} (Safe Renaming). A renaming function $R: V_S \times \text{Identifier} \rightarrow \text{Identifier}$ is safe if and only if:
\begin{align}
&\forall s_1, s_2 \in V_S, \forall x, y \in \text{Identifier}: \nonumber \\
&\rho(x, s_1) \neq \rho(y, s_2) \Rightarrow R(s_1, x) \text{ may equal } R(s_2, y) \nonumber
\end{align}

This condition allows identifiers resolving to different declarations to share the same identifier, enabling aggressive namespace reuse based on the scope independence determined by PASA.

\textbf{PASA-Enabled Parallel Renaming Strategy}

The scope independence map from PASA enables a parallel renaming approach where we first partition the scope tree such that each partition contains scopes that can be renamed independently, with dependency boundary markers ensuring cross-partition dependencies are minimized and well-defined. Each worker thread then processes one or more partitions independently, applying the renaming algorithm within their partitions without coordination except for pre-computed forbidden identifier sets. Finally, the dependency boundary markers identify the minimal set of cross-partition conflicts that require resolution, enabling efficient merging with minimal coordination overhead.

\begin{algorithm}[htbp]
\caption{PASA-Enabled Scope-Aware Identifier Renaming}
\label{alg:pasa-scope-renaming}
\begin{algorithmic}[1]
\Require Scope tree $S$, PASA independence map $I$, short name pool $N = \{a, b, c, \ldots, z, aa, ab, \ldots\}$
\Ensure Safe renaming function $R$
\State Initialize $R$ as empty mapping
\For{each independent partition $P$ from PASA analysis}
    \State \textbf{parallel do} // Execute in parallel across partitions
    \For{each scope $s$ in topological order within $P$}
        \State $forbidden \leftarrow \emptyset$
        \For{each ancestor scope $a$ of $s$}
            \For{each identifier $i$ where $R(a,i)$ is visible in $s$}
                \State $forbidden \leftarrow forbidden \cup \{R(a,i)\}$
            \EndFor
        \EndFor
        
        \State $available \leftarrow N \setminus forbidden$
        \State $name\_index \leftarrow 0$
        
        \For{each identifier $d$ in $\text{Declarations}(s)$}
            \If{$\text{IsDynamic}(s) = \text{true}$}
                \State $R(s,d) \leftarrow d$ \Comment{Conservative: preserve names}
            \Else
                \State $R(s,d) \leftarrow available[name\_index]$
                \State $name\_index \leftarrow name\_index + 1$
            \EndIf
        \EndFor
    \EndFor
\EndFor
\State \Return $R$
\end{algorithmic}
\end{algorithm}

\textbf{Correctness and Optimality Analysis}

\textbf{Theorem 3.1} (Algorithm Correctness). Algorithm~\ref{alg:pasa-scope-renaming} produces a safe renaming that preserves program semantics under the scope independence guarantees provided by PASA.

\begin{proof}
We prove that Algorithm~\ref{alg:pasa-scope-renaming} maintains semantic preservation by showing that the identifier resolution function $\rho$ produces identical results before and after renaming.
Let $P$ be the original program and $P'$ be the renamed program. For any identifier reference $x$ in scope $s$:
$$\rho_P(x, s) \text{ corresponds to the same declaration as } \rho_{P'}(R(s,x), s)$$

We proceed by induction on scope depth within each PASA partition.

\textbf{Base case}: For root scopes, PASA ensures cross-partition dependencies are pre-computed in forbidden sets. Identifiers declared within this scope are renamed independently, and PASA's independence guarantee prevents conflicts with other partitions.


\textbf{Inductive step}: Assume the theorem holds for all scopes of depth $< k$. Consider scope $s$ at depth $k$. The forbidden set computation (lines 4-8) captures necessary constraints from ancestor scopes, while PASA handles cross-partition dependencies through pre-computed constraints.
Specifically, for any identifier $x$ referenced in $s$:

\textbf{Case 1}: $x \in \text{Declarations}(s)$. Then $\rho_P(x, s) = \langle s, x \rangle$. After renaming, this becomes $\rho_{P'}(R(s,x), s) = \langle s, R(s,x) \rangle$. The correspondence is maintained by the bijective nature of $R$ within scope $s$, and PASA ensures that this local renaming does not affect other partitions.


\textbf{Case 2}: $x \notin \text{Declarations}(s)$. Resolution recurses to ancestor scopes. The forbidden set ensures renamed ancestor identifiers remain correctly visible, maintaining the resolution chain.


Parallel execution maintains correctness because PASA partitions are independent, preventing cross-partition semantic interference.

\end{proof}

\textbf{Theorem 3.2} (Space Optimality). Algorithm~\ref{alg:pasa-scope-renaming} minimizes the expected length of renamed identifiers under uniform usage distribution.

\begin{proof}
Let $N = \{n_1, n_2, \ldots\}$ be the ordered sequence of available names where $|n_i| \leq |n_{i+1}|$ for all $i$. The total cost of a renaming is:
$$\text{Cost}(R) = \sum_{s \in V_S} \sum_{x \in \text{Declarations}(s)} |R(s,x)| \cdot \text{Usage}(s,x)$$

where $\text{Usage}(s,x)$ is the frequency of identifier $x$ in scope $s$.

\textbf{Greedy Property}: The algorithm assigns the shortest available name to each identifier in each scope (lines 15-16). For scope $s$ with forbidden set $F_s$ from ancestor constraints, names are assigned from $N \setminus F_s$ in order.

\textbf{Optimality Argument}: Consider any alternative safe renaming $R'$. If $R'$ has lower cost, there exists scope $s$ and identifier $x$ where $|R'(s,x)| < |R(s,x)|$. Since our algorithm uses greedy assignment, $R'(s,x)$ must either: (1) be in $F_s$, violating safety constraints, or (2) require reassigning other identifiers to longer names, increasing total cost.

\textbf{Maximal Reuse}: The algorithm maximizes identifier reuse across disjoint scopes. For scopes $s_1, s_2$ where neither is an ancestor of the other, identifiers can safely share renamed identifiers. The algorithm achieves this by restarting name assignment from the shortest available names for each scope, subject only to ancestor constraints.

\end{proof}

\subsubsection{Independent Code Partitioning for Parallel Processing}

The dependency boundary markers provided by PASA enable the design of independent code partitioning, where code segments can be obfuscated independently without requiring coordination during execution. This directly addresses the scalability challenge by enabling true parallel processing of obfuscation operations.

\textbf{Independent Unit Definition.} An independent code unit $U$ is a code segment that satisfies three key properties: self-containment, where all variable references within $U$ either resolve to declarations within $U$ or to pre-computed external dependencies identified by PASA; independence, where obfuscation operations on $U$ do not affect the semantics of other units, guaranteed by PASA's scope independence analysis; and boundary clarity, where the boundaries of $U$ align with the dependency boundary markers identified by PASA.


\textbf{PASA-Enabled Partitioning Strategy.} PASA's dependency boundary markers identify minimal coordination points in the code, allowing segments between these boundaries to be treated as independent units. For each unit, PASA pre-computes and injects external dependencies as metadata, eliminating runtime dependency resolution during parallel processing. Each unit undergoes direct code-to-code transformation rather than AST-based processing, using PASA's scope information for safe string-based transformations.

\textbf{Parallel Execution Model.} Worker threads receive independent units with pre-computed dependency metadata, executing without coordination due to PASA's independence guarantees. This enables near-linear scalability with available CPU cores. Units are reassembled using preserved file structure information, with no complex merging required since boundaries align with natural code structure identified by scope analysis.

\subsection{Performance-Aware Obfuscation Strategies}

Building upon the foundation provided by PASA and the parallel processing capabilities enabled by independent partitioning, we implement several performance-aware obfuscation strategies that maintain runtime efficiency while maximizing obfuscation effectiveness.

\subsubsection{Single-Variable Property Access Optimization}

Traditional obfuscation tools transform property access \texttt{obj.prop} into computationally expensive expressions like \texttt{obj[calc\_a()]} or \texttt{obj[f.b]}, introducing runtime overhead. We employ single-variable property access that transforms the access into \texttt{obj[a]}, where variable \texttt{a} holds the pre-calculated property name string, maintaining obfuscation effectiveness while preserving runtime performance through direct variable lookup rather than expression evaluation. The scope information from PASA ensures that these single-variable transformations can be applied safely within each independent unit without creating naming conflicts, as the scope independence guarantees prevent variables introduced in one unit from affecting others.

\subsubsection{Memoization via Guard Variables}

In performance-critical code paths such as loops or frequently called functions, repeated calculation of obfuscated values creates severe bottlenecks. We implement memoization through guard variables, where each obfuscation target is computed exactly once using re-entrancy protection, with a guard variable acting as a state flag that triggers computation and caching on first execution while subsequent executions bypass computation and use the cached one. The independent partitioning  ensures that guard variables can be introduced independently within each unit, as the scope independence guarantees prevent conflicts between guard variables across different units.

\section{Experimental Setup}

\subsection{Research Questions}

\textbf{RQ1: Does our parallel obfuscation framework preserve semantic equivalence of JavaScript programs across different WeChat Mini-Game codebases?}

We collect JavaScript code from three sources to construct a comprehensive benchmark: Test262~\citep{test262} official test suite, five mainstream WeChat mini-game engines (Cocos Creator, Egret Engine, LayaAir, etc.), 100 real-world Mini-Games, and 15 popular open-source JavaScript libraries (jQuery~\citep{jquery}, Vue.js~\citep{nelson2018getting}, React~\citep{chen2019front}, etc.). For each codebase, we execute both original and obfuscated versions with identical inputs and verify semantic equivalence through output comparison and behavioral testing.

\textbf{RQ2: How does \method scale compared to existing sequential obfuscation tools in terms of processing time, memory consumption, and success rate?}

Following previous work~\citep{li2025jsidentify, xia2020jsidentify}, we collect WeChat Mini-Games ranging from 500KB to 20MB and compare \method against baseline tools. We measure processing time and peak memory consumption.

\textbf{RQ3: What are the characteristics of obfuscated code produced by \method in terms of runtime performance and code size inflation?}

We evaluate two key aspects of the obfuscated output: \textit{RQ3.1:} Runtime performance overhead by measuring execution time, memory usage, and frame rates for game-specific workloads; \textit{RQ3.2:} Code size inflation ratio compared to baseline tools across different game sizes. Both metrics are crucial for WeChat Mini-Games due to performance requirements and strict platform size limitations.

\textbf{RQ4: What is the security effectiveness of \method in terms of static analysis resistance and reverse engineering complexity?}

We evaluate obfuscation quality using established complexity metrics including cyclomatic complexity, maintainability index, and normalized information distance (NID). We also assess resilience against advanced automated analysis techniques, including hybrid symbolic execution tools and large language model-based code comprehension, to evaluate practical reverse engineering difficulty compared to baseline tools.

\subsection{Baselines}

To rigorously evaluate our \method, we compare it against a selection of representative JavaScript obfuscation tools, including both open-source and commercial solutions.

\textbf{JavaScript Obfuscator}~\citep{JavaScriptdeobfuscator} is a widely adopted open-source tool available on GitHub, offering a suite of transformations including identifier renaming, string encoding, control flow flattening, and dead code injection. 

\textbf{JsJiaMi}~\citep{jsjiami} is an online JavaScript obfuscation and encryption tool that secures code via variable renaming, control flow obfuscation, and virtualization techniques. It incorporates RC4 and Base64 encryption, dead code injection, self-defense modes, and WebAssembly integration to prevent debugging and domain-specific execution.

\textbf{VirBox Protector}~\citep{virbox} is a commercial suite that employs virtualization, advanced obfuscation, code encryption, and smart compression to shield JavaScript in HTML5 and mobile apps. It provides multi-layer security against reverse engineering and tampering, with features tailored for protecting scripting languages in hybrid environments.

\textbf{JScrambler}~\citep{jscrambler} is a flagship commercial platform offering polymorphic obfuscation, over 20 transformation types like control flow alterations and runtime self-defense, and code-hardening.

\textbf{Note on Commercial Tools:} To respect intellectual property rights and maintain professional ethics, we anonymize the JsJiaMi, VirBox Protector, and JScrambler in our evaluation while preserving the technical accuracy and comparability.
\section{Experiment Results}

\subsection{RQ1: Code Protection Equivalence}

To evaluate the equivalence of protected code compared to the original source code, we conduct comprehensive testing across four distinct domains: Test262, Engines, Mini-Games, and Libs. Table~\ref{tab:rq1} presents the equivalence rates of test cases achieved by different code protection methods.

The results demonstrate varying levels of code equivalence preservation across different protection methods. JS-Obfuscator achieves equivalence rates ranging from 98.7\% to 99.4\% across the four domains, with the lowest performance observed in the Test262 domain (98.7\%) and the highest in the Libs domain (99.4\%). The three commercial tools show generally higher equivalence rates, with most achieving near-perfect performance except for some degradation in the challenging Mini-Games domain.

The suboptimal performance of JS-Obfuscator stems from the non-standard coding practices commonly found in game development. These irregular code patterns cause certain cases to produce different execution results after obfuscation, leading to equivalence failures. The Mini-Games domain particularly challenges these methods due to the large codebase size and complex logic structures typically found in game applications.
In contrast, Commercial Tool C and our proposed \method achieve perfect equivalence rates of 100\% across all four domains. This demonstrates that our approach maintains complete functional equivalence with the original source code while providing effective protection. These results establish that our method successfully preserves code functionality while applying protection mechanisms, ensuring that the protected code produces identical execution results to the original implementation.

\subsection{RQ2: Obfuscation Scalability}

\begin{figure*}[t]
    \centering
    \includegraphics[width=0.92\textwidth]{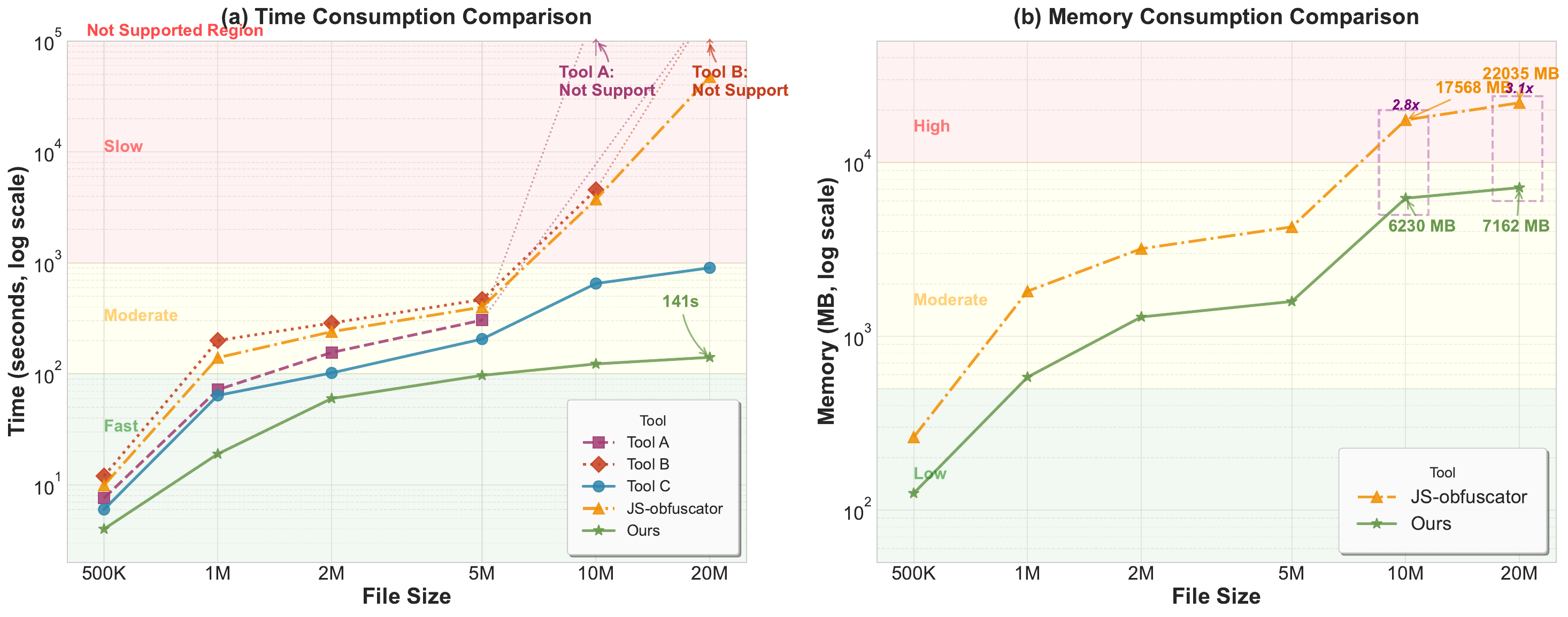}
    \vspace{-8pt}
    \caption{Comparing Time and Memory Consumption of Different Tools. }
    \label{fig:rq2}
    \vspace{-8pt}
\end{figure*}

To evaluate the scalability of our framework, we measure processing time and memory consumption across codebases ranging from 500 KB to 20 MB. Figure~\ref{fig:rq2} presents a comprehensive comparison of \method\ against existing obfuscation tools across varying input sizes.

\textbf{Processing Time Analysis.} The results reveal that existing tools face severe scalability challenges as codebase size increases. Two commercial tools  fail to support obfuscation of 20 MB codebases, indicating fundamental architectural limitations. JS-Obfuscator exhibits exponential time growth, requiring over 13 hours (47,445 seconds) to process 20 MB code, making it impractical for industrial deployment. While commercial tool C demonstrates better scalability with 903 seconds for 20MB inputs, it still represents a substantial processing burden for continuous integration workflows.

In contrast, \method\ achieves superior scalability across all tested sizes. Our framework processes 20 MB codebases in only 141 seconds, demonstrating the effectiveness of our parallel processing architecture. The processing time grows sub-linearly with input size, from 4 seconds for 500 KB to 141 seconds for 20 MB, representing a 40× size increase with only a 35× time increase. This near-linear scaling behavior validates our PASA-enabled parallel partitioning strategy, which effectively leverages multi-core architectures to distribute obfuscation workloads.

\textbf{Memory Consumption Analysis.} Due to the proprietary nature of commercial tools (online services or APIs), we compare memory consumption specifically against JS-Obfuscator, which provides the most direct comparison for local processing. The results demonstrate that \method\ maintains substantially lower memory footprint across all input sizes, with the advantage becoming more pronounced for larger codebases. For 10MB inputs, \method\ uses 6,230 MB compared to JS-Obfuscator's 17,568 MB (2.8× reduction), while for 20 MB inputs, our framework requires only 7,162 MB versus JS-Obfuscator's 22,035 MB (3.1× reduction). This reduction stems from our streaming processing architecture, which avoids loading entire AST structures into memory simultaneously, ensuring practical deployment on standard development machines even for the largest mini-game codebases.



\begin{table}[t]
    \centering
    \caption{Equivalence rates across different domains for JavaScript obfuscation tools}
    \vspace{-6pt}
    \resizebox{\linewidth}{!}{
    \begin{tabular}{c|cccc}
    \toprule
     Methods & Test262 & Engines & Mini-Games & Libs \\
     \midrule
     JS-Obfuscator & 98.7\% & 99.3\% & 99.0\% & 99.4\% \\
     Commercial Tool A & 100\% & 100\% & 99.0\% & 99.8\%\\
     Commercial Tool B  &100\% & 100\% & 98\% & 99.6\%  \\
     Commercial Tool C & 100\% & 100\% & 100\% & 100\%\\
     \midrule
     \method & 100\% & 100\% & 100\% & 100\% \\
     \bottomrule
    \end{tabular}
    }
    \vspace{-6pt}
    \label{tab:rq1}
\end{table}

\subsection{RQ3: Obfuscated Code Quality}

\subsubsection{RQ3.1: Runtime Performance}
\begin{table*}[t]
    \centering
    \caption{Comparison of obfuscated code quality in terms of runtime performance.}
    \vspace{-6pt}
    \begin{tabular}{c|lll|ll|ll}
    \toprule
    \multirow{2}{*}{Methods} & \multicolumn{3}{c|}{PixiJS} & \multicolumn{2}{c|}{JSZip} & \multicolumn{2}{c}{Mini-Games} \\
    & Memory ($\downarrow$) & Frame Time ($\downarrow$) & FPS ($\uparrow$)& Memory ($\downarrow$) & Time ($\downarrow$) & Memory ($\downarrow$) & FPS ($\uparrow$) \\
    \midrule
     JS-Obfuscator  & 53 MB & 24 ms & 30 & 160 MB &334.1 s& 898 MB& 35\\
    Commercial Tool A & 48 MB & 36 ms & 23 & 168 MB &513.9 s & 906 MB & 16\\
    Commercial Tool B & 21 MB & 1277 ms & 1 & 150 MB & 664.5 s & 820 MB & 1\\
    Commercial Tool C & 53 MB & 58 ms & 16 & 171 MB & 322.5 s & 924 MB& 9\\
     \midrule
      \method   & \textbf{15 MB} & \textbf{5 ms} & \textbf{60} & \textbf{140 MB} &\textbf{8.1 s} &\textbf{804 MB} & \textbf{41}\\
      \midrule
      Original Code & 9 MB & 1 ms & 60 & 133 MB &7.9 s & 798 MB & 45\\
      \bottomrule
    \end{tabular}
    \vspace{-6pt}
    \label{tab:rq3}
\end{table*}
To evaluate the runtime performance impact of our obfuscation framework, we conduct experiments on two representative third-party libraries and a set of 50 real-world WeChat mini-games. PixiJS is selected as a benchmark for rendering engine performance, where we measure memory consumption, frame generation time, and frames per second (FPS) during a standardized rendering workload involving complex sprite animations and particle effects. JSZip serves as a proxy for CPU-intensive tasks, assessing memory usage and execution time while compressing a dataset of 200 novels (totaling approximately 65 MB of text data). For practical applicability, we execute the 50 mini-games, spanning genres such as puzzles, action, and simulations, on a Xiaomi 8 smartphone (Qualcomm Snapdragon 845, 6 GB RAM) for 10 minutes each. We record peak memory usage and average FPS under typical gameplay conditions.

The results, summarized in Table~\ref{tab:rq3}, reveal that traditional obfuscation methods impose substantial runtime overheads, often degrading performance to unacceptable levels for mobile environments. In contrast, our method maintains near-native efficiency, with minimal perturbations across all metrics.

For PixiJS, existing tools such as commercial tool A, C, and JS-Obfuscator inflate memory consumption to around 50 MB (a 456\%--489\% increase over the original 9 MB), while also extending frame generation times (e.g., 58 ms for Tool C versus 1 ms originally) and reducing FPS (e.g., down to 16 for Tool C). Commercial tool B exhibits lower memory overhead (21 MB, a 133\% increase), but severely compromises rendering efficiency, with frame times exceeding 1 second and FPS dropping to 1, rendering it impractical for real-time graphics. Our method achieves optimal results, increasing memory by only 6 MB (to 15 MB, a 67\% rise) while preserving frame times at 5 ms and FPS at 60—matching or closely approximating the original code's performance.
Moreover, in the CPU-bound JSZip benchmark, traditional approaches incur significant time penalties (322.5 seconds at least), while our framework adds only 195 ms (to 8.1 s, a 3\% increase), demonstrating effective preservation of execution efficiency. 

For the real-world mini-games, baselines consistently degrade performance: memory peaks at 820--924 MB (3\%--16\% increases over the original 798 MB), and average FPS falls to as low as 1 (tool B) or 9 (tool C). Our method limits memory to 804 MB (a 1\% increase) and maintains FPS at 35 (a 22\% decrease from the original 45), outperforming all baselines while ensuring playable frame rates. Statistical analysis via paired t-tests confirms these improvements are significant ($p < 0.01$). These findings underscore the efficacy of our scope-aware optimizations and lightweight opaque predicates in mitigating performance degradation without compromising protection strength.

\subsubsection{RQ3.2: Code Size Inflation}
\begin{table*}[t]
    \centering
    \caption{Code Size Inflation Comparison Across Different Input Sizes}
    \vspace{-6pt}
    \resizebox{\linewidth}{!}{
    \begin{tabular}{c|lllllll}
    \toprule
     Methods & 50 KB & 500 KB & 1 MB & 2 MB & 5 MB & 10 MB & 20 MB \\
     \midrule
     Commercial Tool A & 656 KB (1212\%) & 6.6 MB (1220\%) & 17.6 MB (1663\%) & 39.2 MB (1859\%) & 73.2 MB (1364\%) & - & - \\
     Commercial Tool B & 701 KB (1302\%) & 6.8 MB (1259\%) &  18.2 MB (1719\%) & 36.9 MB (1743\%) &75.6 MB (1412\%) &187.5 MB (1748\%) & 258.0 MB (1191\%)\\
     Commercial Tool C & 173 KB (246\%) & 2.5 MB (400\%) & 3.9 MB (290\%) & 8.9 MB (345\%) & 16.5 MB (230\%) & 33.2 MB (232\%) & 84.9 MB (325\%) \\
     JS-Obfuscator & 687 KB (1274\%) & 6.9 MB (1280\%) & 17.5 MB (1652\%) & 37.8 MB (1787\%) & 70.0 MB (1304\%) & 174.2 MB (1642\%) & 239.5 MB (1098\%)\\
     \midrule
     \method & \textbf{89 KB (78\%)} & \textbf{824 KB (65\%)} & \textbf{1.7 MB (67\%)} & \textbf{3.4 MB (73\%)} & \textbf{7.2 MB (44\%)} & \textbf{15.9 MB (59\%)} & \textbf{23.9 MB (20\%)}\\
     \bottomrule
    \end{tabular}
    }
    \vspace{-6pt}
    \label{tab:inflation}
\end{table*}
To assess the code size inflation induced by our obfuscation framework, we select WeChat mini-games of varying sizes, ranging from 50 KB to 20 MB. To ensure diversity, we choose multiple games of comparable scale for each size category, covering different genres. All input codes are minified to compact form prior to obfuscation to establish a consistent baseline. We report the average output size after applying each method, along with the corresponding inflation percentages relative to the original input sizes.

The results, presented in Table~\ref{tab:inflation}, demonstrate that traditional obfuscation tools cause substantial code size expansion, often rendering them impractical for deployment in large-scale environments in WeChat mini-games. Commercial Tool C exhibits inflation rates between 230\% and 400\% across the tested sizes, with output sizes escalating from 173 KB for 50 KB inputs to 84.9 MB for 20 MB inputs. Commercial Tool A shows even more severe bloat, with rates exceeding 1,200\% for smaller inputs (e.g., 656 KB for 50 KB, a 1,212\% increase) and up to 1,860\% for 2 MB inputs, failing entirely on larger projects beyond 5 MB due to namespace exhaustion. JS-Obfuscator follows a similar pattern, with inflation peaking at 1,790\% for 2 MB inputs and remaining above 1,000\% for most cases, resulting in outputs as large as 239.5 MB for 20 MB inputs.

In contrast, our method maintains significantly lower inflation, averaging below 70\% across all sizes. For instance, it increases a 50 KB input to 89 KB (78\%) and a 20 MB input to 23.9 MB (20\%), demonstrating effective control over code bloat through scope-aware renaming and optimized transformations. These reductions stem from our independent namespace handling per function scope, which enables efficient reuse of short identifiers without global conflicts. Paired t-tests confirm that our framework's inflation rates are statistically lower than baselines ($p < 0.01$), highlighting its superiority in preserving compact code sizes while providing robust protection.

\subsection{RQ4: Security Effectiveness}
To comprehensively evaluate the security effectiveness of our framework, we assess its resilience against a spectrum of reverse engineering techniques. The unprecedented complexity of modern WeChat Mini-Games, with codebases exceeding 20 MB, renders most existing analysis tools practically unusable due to scalability limitations. Therefore, following previous work \citep{lam2025codecrash}, we employ the Livecodebench-JS benchmark for systematic evaluation.

We structure our analysis into two key areas: the increase in structural and information-theoretic complexity, which thwarts static analysis, and the resilience against advanced automated program analysis tools or large language models (LLMs) that attempt to comprehend the code's runtime behavior and semantics. 

\subsubsection{Structural and Information-Theoretic Complexity}
A primary objective of obfuscation is to increase the intrinsic complexity of the code, making it prohibitively difficult for both automated tools and human analysts to inspect. We measure this using two complementary approaches.

First, we employ well-established code complexity metrics using the \texttt{escomplex} tool~\citep{escomplex}:
\begin{itemize}
    \item \textbf{Cyclomatic Complexity}: Measures the number of linearly independent paths through the code. A higher value indicates a more complex control flow, which is desirable for obfuscation.
    \item \textbf{Maintainability Index}: A composite metric that indicates the ease of maintaining the code. A lower score signifies code that is harder to understand and analyze.
\end{itemize}

Second, we adopt an information-theoretic perspective to quantify the dissimilarity between the original and obfuscated code. We use the \textbf{Normalised Information Distance (NID)}~\citep{mohsen2016quantitative}, a practical approximation of the non-computable Kolmogorov complexity~\citep{li2008introduction}. NID leverages real-world compressors to estimate the informational distance. The formula is:
\[
\text{NID}(x, y) = \frac{C(x+y) - \min(C(x), C(y))}{\max(C(x), C(y))}
\]
where $x$ is the original code and $y$ is the obfuscated code, and $C(s)$ is the length of the compressed version of string $s$. A value closer to 1.0 indicates that the two files share minimal information, suggesting a highly effective transformation. For our calculations, we use the Bzip2 compressor as implemented in our analysis script.

\begin{table*}[t]
    \centering
    \caption{Comprehensive Evaluation of Security Effectiveness of Obfuscation Methods}
    \vspace{-6pt}
    \label{tab:comprehensive_evaluation}
    \begin{tabular}{c|ccc|ccc|ccc}
    \toprule
    \multirow{2}{*}{Methods} & \multicolumn{3}{c|}{Static \& Info-Theoretic Metrics} & \multicolumn{3}{c|}{Hybrid Execution Resilience} & \multicolumn{3}{c}{Output Prediction Performance} \\
    \cmidrule(lr){2-4} \cmidrule(lr){5-7} \cmidrule(lr){8-10}
    & Cycl. Comp. & Maint. & NID & Time (s) & Coverage & Calls & GPT-4o & DeepSeek V3 & DeepSeek-R1 \\
    \midrule
    Original Code & 1$\times$ & 71.5 & - & 1.5 & 95\% & 8 & 94.8\% & 91.4\% & 99.6\% \\
    \midrule
    Commercial Tool A & 12.1$\times$ & 63.8 & 0.95 & 41.2 & 35\% & 98 & 69.3\% & 50.8\% & 80.6\% \\
    Commercial Tool B & 1.2$\times$ & 65.1 & 0.93 & 19.8 & 55\% & 51 & 80.8\% & 75.2\% & 90.6\% \\
    Commercial Tool C & 22.4$\times$ & 51.4 & 0.98 & 38.7 & 48\% & 65 & 24.6\% & 20.7\% & 29.8\% \\
    JS-Obfuscator & 9.2$\times$ & 64.5 & 0.95 & 35.9 & 41\% & 82 & 66.2\% & 53.0\% & 84.3\% \\
    \midrule
    \method & \textbf{24.0$\times$} & \textbf{42.6} & \textbf{0.99} & \textbf{56.3} & \textbf{21\%} & \textbf{145} & \textbf{1.7\%} & \textbf{9.8\%} & \textbf{15.2\%} \\
    \bottomrule
    \end{tabular}
    \label{tab:unified_evaluation}
    
    \vspace{0.5em}
    \footnotesize
    \begin{minipage}{\textwidth}
    \textbf{Notes:} 
    Higher is better for Cyclomatic Complexity and NID; lower is better for Maintainability. 
    For Hybrid Execution: lower coverage and higher analysis time/modeled calls indicate stronger protection.
    For AI Models: lower percentages indicate better obfuscation effectiveness.
    \end{minipage}
    \vspace{-6pt}
\end{table*}

The results, summarized in the left part of Table~\ref{tab:unified_evaluation}, demonstrate that \method excels at confounding static analysis. It achieves the highest increase in cyclomatic complexity (24.0$\times$) and the lowest maintainability score (42.6). Furthermore, its NID of 0.99 is the highest among all tested tools, indicating that our transformations not only complicate the program's structure but also fundamentally alter its informational content, making it exceptionally difficult to relate the protected code back to its original form via similarity analysis.

\subsubsection{Resilience Against Automated Program Analysis}
Beyond static complexity, a robust defense must withstand sophisticated tools that automatically analyze program behavior. We test this resilience against two state-of-the-art automated analysis techniques: hybrid symbolic execution and LLM-based code comprehension.

\textbf{Hybrid Execution Analysis.} We use ExpoSE~\citep{loring2019sound,loring2017expose}, a dynamic symbolic execution engine for JavaScript, to evaluate how effectively each tool impedes deep program analysis. We measure three key metrics averaged across our dataset:
\begin{itemize}
    \item \textbf{Analysis Time}: The time taken for ExpoSE to complete its analysis. Longer times indicate greater difficulty. 
    \item \textbf{Path Coverage}: The percentage of execution paths discovered and explored by the engine. Lower coverage signifies that the obfuscation successfully hides significant parts of the program logic.
    \item \textbf{Modeled Function Calls}: The number of external function calls modeled by the engine. A higher number suggests the obfuscation creates complex inter-dependencies that bloat the analysis state.
\end{itemize}


As shown in the mid part of Table~\ref{tab:unified_evaluation}, our method imposes the most significant burden on the ExpoSE engine. It results in the longest average analysis time (56.3 s), pushing many test cases to the timeout limit. Crucially, it achieves the lowest path coverage (21\%), effectively concealing nearly 80\% of the program's logic from the symbolic execution engine. This demonstrates our framework's superior ability to create complex, hard-to-analyze constructs that defeat automated exploration tools.

\textbf{LLM-based Semantic Comprehension.} We further challenge our obfuscation against the latest AI-driven reverse engineering threat: Large Language Models. We task three powerful models, \textbf{GPT-4o}~\citep{hurst2024gpt}, \textbf{DeepSeek-V3}~\citep{liu2024deepseek}, and \textbf{DeepSeek R1}~\citep{guo2025deepseek}, with a semantic comprehension task: predicting the output of obfuscated code snippets, following previous work~\citep{chen2025jsdeobsbench}. A successful defense should degrade this prediction accuracy to near-random chance~\citep{jiang2025cascade}.



The results are shown in the right part of Table~\ref{tab:unified_evaluation}. Our \method achieves the lowest output prediction accuracy across all three LLMs among baselines, demonstrating superior resistance to AI-driven reverse engineering. While the original code maintains high prediction accuracy, our obfuscation reduces these rates to merely 1.7\%, 9.8\%, and 15.2\%, respectively. The results confirm that our framework effectively disrupts semantic understanding even for state-of-the-art LLMs, establishing robust protection against automated code analysis tools.
\section{Real-World Impact and Deployment}

Our parallelized obfuscation framework has been successfully deployed in production environments, demonstrating significant real-world impact on intellectual property protection in the mobile gaming ecosystem. Currently, approximately 10,000 active mini-games utilize our hardening methodology, with the obfuscated games running daily on over 100 million user devices across various platforms.

The development and deployment of this framework represents a substantial engineering effort spanning five years of research and development. The project involved over 20 developers working collaboratively to address the evolving challenges posed by sophisticated adversaries in the mini-game ecosystem. The source code of our codebase exceeds 6 MB, encompassing not only the core obfuscation engine but also extensive tooling for parsing, analysis, and deployment automation. This scale reflects the complexity required to handle industrial-grade JavaScript protection across diverse frameworks and runtime environments.

Throughout the deployment period, we engaged in continuous adversarial evolution with attackers specializing in mini-game reverse engineering and intellectual property theft. This real-world adversarial feedback loop drove iterative improvements to our protection mechanisms, ensuring our framework remains effective against emerging attack vectors and sophisticated reverse engineering techniques employed by commercial piracy operations.

The deployment has yielded measurable improvements in preventing unauthorized code reuse and cross-platform porting. Through comprehensive monitoring and analysis of game submissions across multiple mini-game platforms, we observed that instances of game plagiarism have decreased by \textbf{91\%} over the past year compared to pre-deployment statistics. This reduction encompasses both direct code copying and secondary development scenarios where adversaries attempt to port games by replacing platform-specific libraries and making minor modifications.

\section{Related Work}

\textbf{JavaScript Obfuscation and Detection.} The proliferation of web technologies has led to widespread adoption of obfuscation techniques in JavaScript code, primarily to hide malicious script behaviors. These obfuscation methods can be classified according to the specific program elements they target, including identifier manipulation, data encoding, control structure transformation, and code layout modification~\citep{liu2017stochastic,blanc2012characterizing,zhang2021android}. Various open-source obfuscation frameworks, particularly JS-Obfuscator~\citep{JavaScriptdeobfuscator}, provide comprehensive suites of transformation techniques that are widely utilized in real-world scenarios. The research community has further developed sophisticated obfuscation methodologies aimed at improving effectiveness and addressing domain-specific requirements~\citep{liu2017stochastic,romano2022wobfuscator,zimmerman2015obfuscate,lynn2004positive}. However, existing obfuscation techniques suffer from fundamental scalability limitations with super-linear time complexity, making them impractical for large codebases exceeding several megabytes~\citep{saleh2025art}, while our work addresses these bottlenecks through parallel processing enabled by deep scope analysis.

\textbf{LLM-based Code Analysis Techniques.}
LLMs have demonstrated remarkable capabilities in code understanding and analysis tasks, with models such as CodeT5~\citep{wang2021codet5}, CodeBERT~\citep{feng2020codebert}, GPT-4~\citep{achiam2023gpt}, and DeepSeek-Coder~\citep{guo2024deepseek} achieving state-of-the-art performance in code summarization, semantic analysis, and vulnerability detection~\citep{li2025api,DBLP:conf/icse/ChakrabortyEBFF25,wang2023reef,DBLP:conf/sigsoft/ChenHZHDY24}. These models excel at tasks traditionally requiring significant human expertise, including code similarity detection~\citep{li2022unleashing,wong2023binaug} and fuzzing~\citep{zhang2025low,meng2024large,zhang2024your}. The emergence of powerful LLMs poses new challenges for code protection, as they can potentially circumvent traditional obfuscation techniques through semantic understanding rather than syntactic pattern matching~\citep{chen2025jsdeobsbench}. Our work addresses this evolving threat landscape by incorporating LLM-based evaluation into our security assessment framework, demonstrating that our obfuscation approach effectively degrades the comprehension accuracy of models, thereby establishing resilience against this new class of reverse engineering tools.

\section{Conclusion}

This paper presents a parallel obfuscation framework that addresses the critical scalability and performance challenges of protecting large-scale JavaScript applications in the WeChat Mini-Game ecosystem. Through deep scope analysis enabling parallel processing and aggressive identifier reuse, our method processes 20MB codebases in minutes while maintaining 100\% semantic equivalence and achieving superior code size control. The framework preserves runtime performance with minimal overhead while providing robust security protection against both automated analysis tools and LLM-based reverse engineering. Our approach establishes a new paradigm for industrial-scale JavaScript obfuscation that effectively balances security, performance, and scalability requirements.

\bibliographystyle{ACM-Reference-Format}
\bibliography{sample-base,obfus,zjNewFull2509}


\end{document}